\documentclass[twocolumn,showpacs,preprintnumbers,amsmath,amssymb]{revtex4}

\usepackage{graphicx}
\input epsf
\input{epsf.sty}

\begin{document}


\title{New High Field State of Flux Line Lattice in Unconventional
Superconductor CeCoIn$_5$}

\author{T.~Watanabe$^1$, Y.~Kasahara$^1$, K.~Izawa$^1$,
T.~Sakakibara$^1$, C.J. van der Beek$^2$, T.~Hanaguri$^3$,
H.~Shishido$^4$, R.~Settai$^4$, Y.~Onuki$^4$, and Y.~Matsuda$^1$}
\affiliation{$^1$Institute for Solid State Physics, University of
Tokyo, Kashiwanoha, Kashiwa, Chiba 277-8581, Japan}%
\affiliation{$^2$Laboratoire des Solides Irradi\'es, CNRS-UMR 7642,
Ecole Polytechnique, 91128 Palaiseau, France}%
\affiliation{$^3$Department of Advanced Materials Science, University
of Tokyo, Kashiwanoha, Kashiwa, Chiba 277-8581, Japan}%
\affiliation{$^4$Graduate School of Science, Osaka University,
Toyonaka, Osaka, 560-0043 Japan}%


\begin{abstract}

Ultrasound velocity measurements of the
unconventional superconductor
CeCoIn$_5$ with extremely large Pauli paramagnetic susceptibility
reveal an unusual structural transformation of the flux line lattice
(FLL) in the vicinity of the upper critical field.  The transition field
coincides with that at which heat capacity measurements reveal a second
order phase transition. The lowering of the sound velocity at the transition
is consistent with the collapse of the FLL tilt modulus and a crossover to
quasi two-dimensional FLL pinning. These results provide a strong evidence
that the high field state is
the Fulde-Ferrel-Larkin-Ovchinikov phase,  in which the order parameter is
spatially modulated and has planar nodes
aligned perpendicular to the vortices. 

\end{abstract}


\maketitle

In almost all superconductors, once the energy gap in the electronic
spectrum opens at the critical temperature $T_{c}$, only the gap
amplitude, but not the shape and symmetry around the Fermi
surface, changes within the SC state.  The only exceptions
to this well-known robustness of the SC gap symmetry have, until now,
been reported for UPt$_3$\cite{upt3}, Sr$_2$RuO$_4$\cite{maeno},
PrOs$_4$Sb$_{12}$\cite{sku}, as well as for superfluid $^3$He.  In
these materials, multiple SC phases with different symmetries
manifest themselves below $T_{c}$.
Moreover, in these superconductors the SC order parameter possesses a
multiplicity, with 
a near-degeneracy of order parameters with
different symmetries.  Tuning the pairing interaction
by an external perturbation such as a magnetic field then causes a
SC state of given symmetry to undergo a transition to a different SC
state.

On the other hand, in the early 1960's, Fulde and Ferrell and Larkin
and Ovchinikov (FFLO) have developed theories\cite{fflo}, different from
the above-mentioned mechanism, for a novel SC phase with a different
SC gap function.  Generally, in spin singlet superconductors,
superconductivity is suppressed by a magnetic field as a consequence
of its coupling to the conduction electron spins (Pauli
paramagnetism) or to the orbital angular momentum (vortices), both
of which break up the Cooper pairs.  The FFLO phase appears when Pauli
pair-breaking dominates over the orbital
effect\cite{gg,tachiki,adachi,maki,shimahara}.
In the FFLO state, pair-breaking arising from the
Pauli effect is reduced by forming a new pairing state ({\boldmath
$k$}$\uparrow$,  {\boldmath $-k+q$}$\downarrow$) with
$\mid${\boldmath $q$}$\mid$  $\sim 2 \mu_B H / \hbar v_F $ ($v_F$ is the
Fermi velocity) between exchange-split parts of the Fermi surface,
instead of
({\boldmath $k$}$\uparrow$,  {\boldmath $-k$}$\downarrow$)-pairing in
ordinary superconductors.   As a result, a new SC state with
different order parameter appears in the vicinity of the upper
critical field $H_{c2}$. One of the most fascinating aspects of the FFLO
state is that Cooper pairs have a drift velocity in the
direction of the applied field, and develop a spatially modulated
order parameter and spin polarization  with a wave length of the
order of $2 \pi/|{\bf q}|$. 
In spite of enormous efforts to find the FFLO state, it has never
been observed in conventional superconductors.  The question of
observing
the FFLO phase in an unconventional superconductor has only been
addressed more recently. In the last decade, heavy fermion
superconductors
CeRu$_2$ and UPd$_2$Al$_3$ have been proposed as candidates for the
observation of the FFLO state, but subsequent research has
called the interpretation of the data for these
materials in terms of an FFLO state into question\cite{gloos}.

Recently it was reported that CeCoIn$_5$ is a new type of heavy
fermion superconductor with quasi two dimensional (2D) electronic
structure and an effective electron mass $m^{*} \approx 100
m_{e}$\cite{pet}.  CeCoIn$_5$ has the highest transition
temperature ($T_c$=2.3~K) of all heavy fermion superconductors
discovered until now. Subsequent measurements have identified
CeCoIn$_5$ as an unconventional superconductor with, most likely,
$d$-wave gap symmetry\cite{mov,izawa}.  Very recent reports of
heat capacity measurements in field parallel to the $ab$--plane
of CeCoIn$_5$ have raised great interest, because they possibly point
to the occurence of a FFLO phase\cite{radovan,bianch}.  The phase
transition from SC to normal metal at the upper critical field
parallel to $ab$--plane, $H_{c2}^{\parallel}$, is of first order
below approximately 1.3~K,  in contrast to the second order
transition in other superconductors\cite{izawa,tayama,bianch,radovan}.
A second order phase transition line,
which branches from the first order $H_{c2}^{\parallel}$-line and
decreases with decreasing $T$,  was observed within the SC state
below 0.35~K\cite{bianch}.   CeCoIn$_5$  satisfies the requirements for the
formation of FFLO state.  First, it is in the very clean regime, {\em
i.e.} the quasiparticle mean free path is much larger than the coherence
length $\xi$.  Second, the Ginzburg-Landau parameter
$\kappa=\lambda/\xi \gg 10$, with $\lambda$ the penetration length,
is very large.
Thirdly, $H_{c2}^{\parallel}(T=0)$ has an extremely high value of
11.9~T, owing to a very large conduction electron mass and to the two
dimensionality. This situation is favorable for the occurence of the FFLO state
because the Pauli effect may overcome the orbital effect. Fourth, the pairing
symmetry is most likely to be $d$-wave, which greatly extends the stability of
the FFLO state with respect to a conventional superconductor\cite{maki}. On the
basis of some of these arguments, the possibility of a high-field FFLO
state has been evoked\cite{radovan,bianch}.   However, there still is no
corroborated evidence for the FFLO state, because of other possible 
origins for a
second order transition, such as some sort of magnetic transition in
the heavy fermion system.  Therefore, a more detailed experimental
investigation of the vortex state is required in order to shed light on this
subject.\begin{figure}[b]
\begin{center}
\includegraphics[height=90mm]{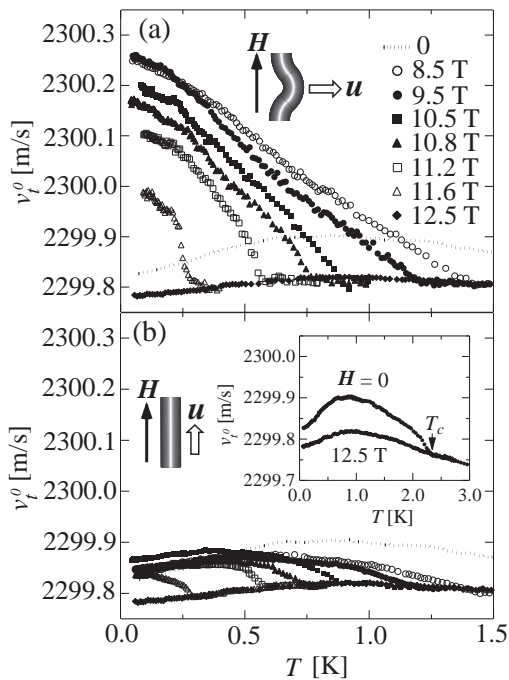}
\caption{The transverse sound velocity $v_t^0$ as a function of
temperature in two different configurations.  (a) the Lorentz mode
${\bf H} \parallel  {\bf k} \parallel $ [100],
${\bf u} \parallel $[010] and (b) the non-Lorentz-mode,
${\bf u} \parallel {\bf H} \parallel $[100],
${\bf k} \parallel $[010]. 
The configuration (a) corresponds to a flux line bending mode.
The inset to Fig.~1(b) shows $v_{t}^{0}$ in zero field and in the
normal state above $H_{c2}^{\parallel}$.}
\end{center}
\end{figure}

The most salient feature of the FFLO state predicted by a
number of theories is that the SC order parameter has planar nodes
aligned perpendicularly to the applied field. In particular, as
pointed out in Ref.~\onlinecite{tachiki},  the flux lines are divided
into segments with distance $\sim 2 \pi /\!\mid \!{\bf q}\!\mid$.
Therefore, in order to establish the existence of a possible FFLO
state, it is particularly important to elucidate the structure of the flux
line lattice (FLL), which in turn is intimately related to the electronic
structure. Here, we present ultrasound velocity measurements which provide
direct information on the FLL structure in CeCoIn$_5$.  In ultrasound
experiments, sound waves are coupled to the FLL when the latter is
pinned by crystal lattice defects. The ensuing modified sound
dispersion allows one to extract detailed information about the
FLL-crystal coupling. We find a distinct
anomaly of the sound velocity, the nature and magnitude of which
provides strong evidence for a vortex state with modulated
order parameter as predicted for the FFLO phase.

The ultrasound velocity measurements were performed on high quality
single crystals of CeCoIn$_5$ with tetragonal symmetry, grown by the
self-flux method.  Ultrasonic waves, with a frequency of 90~MHz,
were generated by a LiNbO$_3$ transducer glued on the surface of the
crystal.  A pulse echo method with a phase comparison technique was
used for the sound velocity measurements.  The relative change of the
sound velocity was measured by the phase change of the detected
signal.   The resolution of the relative velocity measurements was
about 1 part in 10$^6$.   In all measurements, the magnetic field
${\bf H}$ was applied parallel to the 2D planes
(${\bf H} \parallel$ [100]). The inset to Fig.~1(b) shows $v_t^0$
in zero field, and in the normal state above $H_{c2}^{\parallel}$. In
zero field,  $v_t^0$ shows a steep increase below $T_c$, indicating a
stiffening of the crystal lattice at the superconducting transition. In all
experiments presented here, the crystal lattice response is limited by its
shear modulus, $C_{66}^c$.\begin{figure}[b]
\begin{center}
\includegraphics[height=60mm]{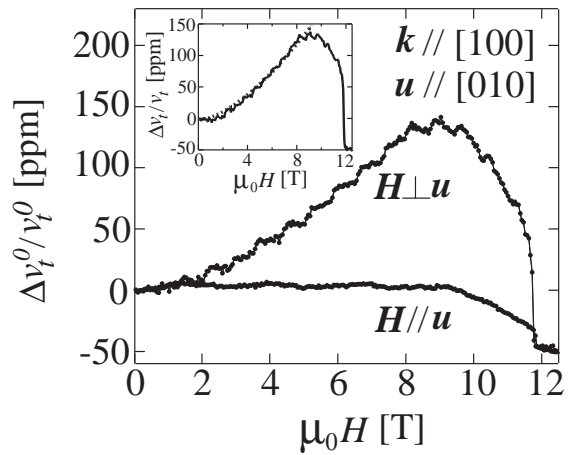}
\caption{The relative shift  of the transverse sound velocity $\Delta
v_{t}^{0}/v_{t}^{0}$ as function of H at 100~mK for the
Lorentz  (${\bf H} \perp {\bf u}$) and the non-Lorentz modes
(${\bf H} \parallel {\bf u}$).  The inset depicts the difference between
Lorentz and non-Lorentz modes, $\Delta v_{t} = v_{t}^{0} ({\bf u}
\perp {\bf H} ) - v_{t}^{0}({\bf u} \parallel {\bf H})$, which can
be attributed to the contribution of the FLL. The dashed line 
shows a fit to
Eq.~(\ref{eq:H-dep}).  }
\end{center}
\end{figure}

Experiments on the FLL were always carried out in the field-cooled
condition.
Figures 1 (a) and 1(b) display the transverse sound velocity $v_t^0$
as a function of temperature in two different configurations with
respect to the polarization vector ${\bf u}$, the sound propagation
vector ${\bf k}$, and ${\bf H}$. We distinguish
between the Lorentz force mode  (Fig. 1 (a), ${\bf u} \perp {\bf H}$)
where the sound wave couples to the vortices through the
induced Lorentz force ${\bf F}_{L} \approx \lambda^{-2}
({\bf u} \times {\bf H}) \times {\bf B}$, and the
non-Lorentz mode (Fig. 1 (b), ${\bf u} \parallel {\bf H}$) where
the sound wave couples only to the crystal lattice. As shown in
Figs.~1 (a) and (b) and Fig.~2, the transverse sound velocity in the
Lorentz mode, in which flux motion is parallel to the
$ab$-plane,  is strongly enhanced compared to that in the
non-Lorentz mode.  This indicates that the transverse ultrasound
strongly couples to the FLL in the Lorentz mode.  The difference
between the sound velocity in the Lorentz mode and in the non-Lorentz
mode, $\Delta v_{t} = v_{t}^{0} ({\bf u} \perp {\bf H}) - v_{t}^{0}({\bf u}
\parallel {\bf H}$),
can be regarded as being the contribution of the FLL to the sound
velocity. In the Lorentz mode, Fig. 1(a), the transverse polarization of the
sound has the FLL undergo a long-wavelength ( $\!\mid\!{\bf
k}\!\mid\!\lambda \ll 1$ ) bending mode,
limited by the FLL tilt modulus in the limit $\!\mid\!{\bf k}\!\mid
\rightarrow 0$, $c_{44}^{f}(\!\mid\!{\bf k}\!\mid \rightarrow
0) = B^{2}/ \mu_{0}$. The FLL contribution can then be written
as\cite{Dominguez}
\begin{equation}
\Delta v_{t} =  \frac{B^{2}}{\mu_{0}\rho v_{t}}
\frac{1}{1 + (\Phi_{0}B)^{1/2} |{\bf k}|^{2}/ \mu_{0}j_{c}},
\label{eq:H-dep}
\end{equation}
\noindent with $\rho$ the mass density of
the superconducting material, $j_{c}$ its critical current density,
and $\Phi_{0} = h/2e$.

Figure 2 shows the relative shift  of
the transverse sound velocity $\Delta v_{t}^{0}/v_{t}^{0}$ as a
function of field at very low temperature.  The hysteresis arising from the
trapped field was very small.  Below approximately 9~T,  $\Delta
v_{t}^{0}/v_{t}^{0}$ for the non-Lorentz mode is nearly
$H$-independent while $\Delta v_{t}^{0}/v_{t}^{0}$ increases
steeply with $H$ for the Lorentz mode. By fitting the increase
to Eq.~(\ref{eq:H-dep}), we obtain a (field--independent) critical
current density, $j_{c}= 7.0 \times 10^{8} {\rm Am^{-2}}$.
In both modes, $\Delta v_{t}^{0}/v_{t}^{0}$ decreases gradually at
higher field with $H$, followed by a jumplike decrease at $H_{c2}^{\parallel}$.
This sharp drop, which occurs in a very narrow $H$-range $\Delta
H/H_{c2}^{\parallel}<1\%$,  is an indication of the first order
transition, and disappears at high temperature where the transition
becomes second order.

In Figure 3, several curves of the relative shift of the FLL
contribution $\Delta v_{t}/v_{t}$ below $H_{c2}$ are shown as a
function of temperature.  As the temperature is lowered below the transition
temperature in magnetic field $T_{c}(H)$, $\Delta v_{t}/v_{t}$ starts to
increase.  At fields of 9.5~T and higher, $\Delta v_{t}/v_{t}$ 
changes its slope
with a distinct cusp at $T^*$, as marked by the arrows, while at
$H$=8.5~T, no anomaly is observed.  Obviously, $T^*$ separates two
different vortex states with different transverse sound velocities;
$\Delta v_{t}/v_{t}$ below $T^*$ is smaller than the $\Delta
v_{t}/v_{t}$ extrapolated from above $T^*$.  This indicates a
sudden decrease of FLL coupling to the crystal lattice below
$T^*$.  Such an anomaly in the ultrasound velocity was not
observed when the magnetic field was swept, probably because it
occurs near the maximum of $\Delta v_{t}(H)$.

Figure 4 displays the $H-T$ phase diagram below 0.7~K.  In this 
temperature range,
the transition at $H_{c2}$ is of first order as indicated by the
sharp drop of the ultrasound velocity (see Fig.~2). In the same
figure, we plot the $H_{c2}$--line reported in
Ref.~\onlinecite{bianch}. Although
$H_{c2}^{\parallel}$ in our crystal is slightly higher than
$H_{c2}^{\parallel}$ reported in Ref.~\onlinecite{bianch}, $T^*$
seems to coincide well with the reported second order transition line\cite{radovan,bianch}. On the basis of our results, we conclude that
{\em the second order phase transition is characterized by a sudden
change of flux line pinning, due to a structural transition of the FLL.} The
present results definitely rule out the possibility that some kind of
magnetic transition, or a structural phase transition of the crystal lattice,
is at the origin of the transition.  This can be seen by looking at $\Delta
v_{t}^{0}/v_{t}^{0}$ in the non-Lorentz modes (Fig.~1 (b)), which
show {\em no} anomaly at $T^*$.


\begin{figure}[t]
\begin{center}
\includegraphics[height=80mm]{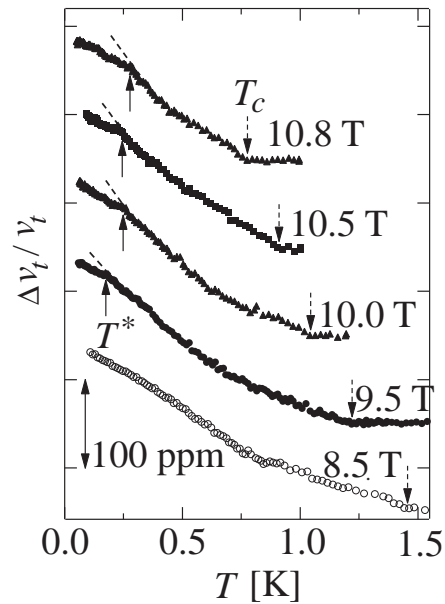}
\caption{The relative shift  of the transverse sound velocity arising
from the contribution of the FLL, $\Delta v_{t}/v_{t}$,  as a
function of temperature, below  the upper critical field
$H_{c2}^{\parallel}$.
$\Delta v_{t}/v_{t}$ is obtained as the difference between  the
non-Lorentz mode and the Lorentz mode sound velocities.  For clarity,
the zero levels of the different curves are vertically shifted.
At $T^*$, marked by arrows,  $\Delta v_{t}/v_{t}$ exhibits a distinct cusp.
The dashed lines are $\Delta v_{t}/v_{t}$ extrapolated from above $T^*$. Dotted arrows
indicate $T_c(H)$.  }
\end{center}
\end{figure}We here discuss  several possible origins for a structural change of
the FLL.  A reduction of the ultrasound velocity with $H$ at the
transition would occur when the vortex lattice melts into a liquid, as observed
in layered high-$T_c$ cuprates and amorphous thin films\cite{Theunissen}. However we can completely rule out this
scenario, because the entropy jump at $T^*$ reported in Ref.
\onlinecite{radovan,bianch} is significantly larger than
that expected for vortex lattice melting. In addition, flux pinning
vanishes at a FLL melting transition, causing $\Delta
v_{t}/v_{t}$ to abruptly drop to zero, at odds with the experimental
observation. A second possibility is a FLL order-disorder transition in the
presence of pinning. However, this should be accompanied by an abrupt
{\em increase} of the measured critical current and of the ultrasound
velocity, whereas the present data show a {\em decrease}. Such an
effect has been discussed in the light of a possible FFLO state
in CeRu$_2$ and  UPd$_2$Al$_3$\cite{Modler}. Thirdly, there
might be a possibility that the nature of FLL pinning changes above
the transition.  For instance, new pinning centers, such as localized spins,
could be induced by the transition.  This effect  may cause the change of the
sound velocity by stiffening the flux lines.  However,  such an
extrinsic origin is quite unlikely because of the absence of an
anomaly in the non-Lorentz mode, as discussed before. Finally, it may be
that the FLL undergoes a symmetry change, {\em e.g.} from hexagonal to
square, with a concomittant lowering of the vortex lattice shear
modulus $c_{66}^{f}$ and increase of the pinning force. Such a
transition has indeed been observed in CeCoIn$_{5}$\cite{Eskildsen},
but at $\mu_{0}H = 0.6$ T, much lower than the fields involved here.

\begin{figure}[t]
\begin{center}
\includegraphics[height=60mm]{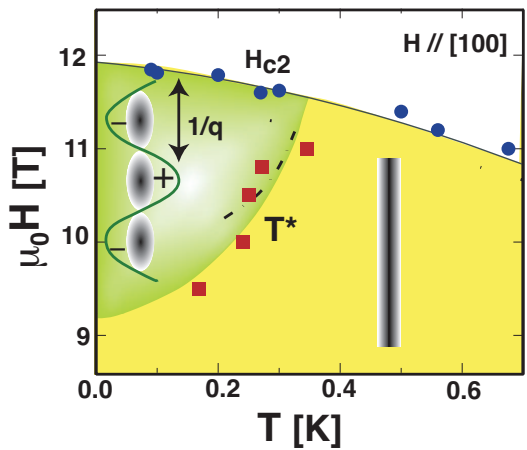}
\caption{Experimental $H-T$ phase diagram for CeCoIn$_5$ below 0.7~K
for (${\bf H}
\parallel$~[100]).
The transition temperature $T^*$, indicated by
 red squares, were obtained from ultrasonic measurements as
indicated by the arrows in Fig.~3.   Blue circles and solid 
line depict the
upper critical field $H_{c2}$ determined by the ultrasonic
experiments. In
this temperature regime,  the transition at $H_{c2}^{\parallel}$
is of first order. The dash-dotted line is the second order
transition line determined by the 
heat capacity
reported in Ref.~\cite{bianch}.  The region shown by green is in the 
FFLO state.  The schematic figures are sketches of
a flux line above and below the transition.    }
\end{center}
\end{figure}
The failure of all the above scenarios leads us to consider the
possibility that the anomaly in the FLL pinning force and the sound velocity
is intimately related to a change of the quasiparticle structure
in the high field SC phase, with a reduced electron
correlation along the direction of the FLL in the high field phase.
Starting from the hypothesis of a transition from a ``continuous''
superconductor to one with an order parameter structure, modulated in
the field direction, with no electronic correlations between layers,
we can understand the modest decrease of $\Delta v_{t}/v_{t}$. From the
critical current density above $T^{*}$, we deduce the pinning strength $W$
of crystalline defects such as small dislocation loops or stacking 
faults. For a
three--dimensional superconductor, this should be evaluated in the
single vortex limit\cite{Wordenweber}: $W =
(\Phi_{0}^{7/2}B^{3/2}j_{c}^{3}/ 4 \pi^{2} \mu_{0} \lambda^{2})^{1/2} =
5 \times 10^{-4}{\rm N^{2}m^{3}}$. At the
transition to the phase with modulated order parameter, the vortex
tilt stiffness in the limit of the {\em small scale distorsions}
caused by pinning collapses. Then, the critical current is given by
the expression for a 2D layer of thickness $d$, $j_{c}^{2D} = W /
\Phi_{0}^{1/2}B^{1/2} c_{66}^{f} d$. At the transition, $j_{c}^{2D}$
must be equal to the experimental value $7 \times 10^{8} {\rm
Am^{-2}}$, an equality, that, given the experimental value of $W$, is satisfied
for a  layer thickness of $4 \times 10^{-8}$ m. This is remarkably
close to the modulation $2 \pi/\!\mid\!{\bf q}\!\!\mid = (m_{e}/m^{*})
( h k_{F} / e B) = 3.5 \times 10^{-8}$ m expected for the order
parameter structure associated with the formation of the FFLO state
proposed by Ref.~\onlinecite{gg,tachiki,adachi,maki,shimahara} (the Fermi
wavevector $k_{F} \approx 8 \times 10^{9} {\rm m^{-1}}$). In the FFLO
state, the
order parameter performs one-dimensional spatial modulations with a
wavelength of the order of $2 \pi /\!\mid\!\!{\bf q}\!\!\mid$, which is
comparable to $\xi$,  along the magnetic field,  forming planar nodes
that are periodically aligned perpendicularly to the flux lines. The
occurrence of planar nodes leads to a segmentation of the flux
lines.  A  schematic figure of this state is shown in Fig.~4.
This field-induced layered structure resembles the vortex
state of layered cuprates and organic superconductors in magnetic
field perpendicular to the conducting planes. The segmentation of vortex lines
to a length $2 \pi /\!\mid\!\!{\bf q}\!\!\mid$ much smaller than the
characteristic ``Larkin length'' $L_{c} = (c_{44}^{f}B/\Phi_{0}W^{1/2})^{2/3}$
on which they are pinned, means that the vortex part of the ultrasound
velocity is now governed by the pinning of the quasi 2D vortex layers
of thickness $2 \pi /\!\!\mid\!{\bf q}\!\!\mid$. Note that a {\em
decrease} of $\Delta v_{t}/v_{t}$ at the transition can {\em only} be
obtained by a true dimensionality change of the FLL at the transition, and
not by a mere softening of the FLL elastic constants.

In summary,  ultrasound velocity measurements of CeCoIn$_5$  in the
vicinity of the upper critical field $H_{c2}$ reveal an unusual
structural transformation of the FLL at the second order phase
transition within the SC state.  This transformation is most likely
characterized by
the collapse of the FLL tilt modulus and a transition to a quasi-2D
state.   These results are a strong indication that the high field
superconducting state is the FFLO phase,
 in which the order parameter is spatially
modulated and has planar nodes aligned perpendicularly to the FLL. 

We thank H.~Adachi, R.~Ikeda, K.~Maki, M.~Nohara,
H.~Shimahara, and M.~Tachiki for stimulating discussions.

\end{document}